\documentclass{article}
\usepackage{spconf,amsmath,graphicx}
\usepackage{multirow}
\usepackage{booktabs}
\usepackage{threeparttable}

\usepackage{color}

\title{Perceptual-assisted Adversarial Adaptation for Choroid Segmentation in optical coherence tomography}
\name{Zhenjie Chai$^{1,2,\dagger}$, Kang Zhou$^{1,2,\dagger}$, Jianlong Yang$^{2\star}$, Yuhui Ma$^{2}$, Zhi Chen$^{3\star}$, Shenghua Gao$^{1}$, Jiang Liu$^{4}$\thanks{Thanks to Ningbo 3315 Innovation team grant.\quad$^\star$Corresponding author: yangjianlong@nimte.ac.cn, Peter459@aliyun.com\quad$^{\dagger}$ Equal contribution}}

\address{
$^{1}$ School of Information Science and Technology, ShanghaiTech
University\\ $^{2}$ Cixi Institute of Biomedical Engineering, \\ Ningbo Institute of Industrial Technology, Chinese Academy of Sciences, Ningbo, China \\$^{3}$Department of Ophthalmology, Fudan University Eye and ENT Hospital, \\Shanghai, China\\ $^{4}$Southern University of Science and Technology, Shenzhen, China }

\newcommand{\etal}{\textit{et~al.}}

\newcommand{\eg}{\textit{e.g.}}
\begin{document}
%\ninept
%
\maketitle
\begin{abstract}
 
Accurate choroid segmentation in optical coherence tomography (OCT) image is vital because the choroid thickness is a major quantitative biomarker of many ocular diseases. Deep learning has shown its superiority in the segmentation of the choroid region but subjects to the performance degeneration caused by the domain discrepancies (\eg , noise level and distribution) among datasets obtained from the OCT devices of different manufacturers. In this paper,  we present an unsupervised perceptual-assisted adversarial adaptation (PAAA) framework for efficiently segmenting the choroid area by narrowing the domain discrepancies between different domains. The adversarial adaptation module in the proposed framework encourages the prediction structure information of the target domain to be similar to that of the source domain. Besides, a perceptual loss is employed for matching their shape information (the curvatures of Bruch's membrane and choroid-sclera interface) which can result in a fine boundary prediction. The results of quantitative experiments show that the proposed PAAA segmentation framework outperforms other state-of-the-art methods.

\end{abstract}

\section{Introduction}
\label{sec:intro}

The choroid in OCT contains abundant blood vessels which are extremely important for the supply of oxygen and nutrients to the outer retina~\cite{Javier2017Open}. Accurate choroid segmentation is a prerequisite for the diagnosis of many ocular diseases. However, manually choroid segmentation by professional ophthalmologists is a heavy and tedious task. An efficient automatic choroid segmentation algorithm is urgently needed.

\begin{figure}[htb]
	\centering
	\includegraphics[width=8.5cm]{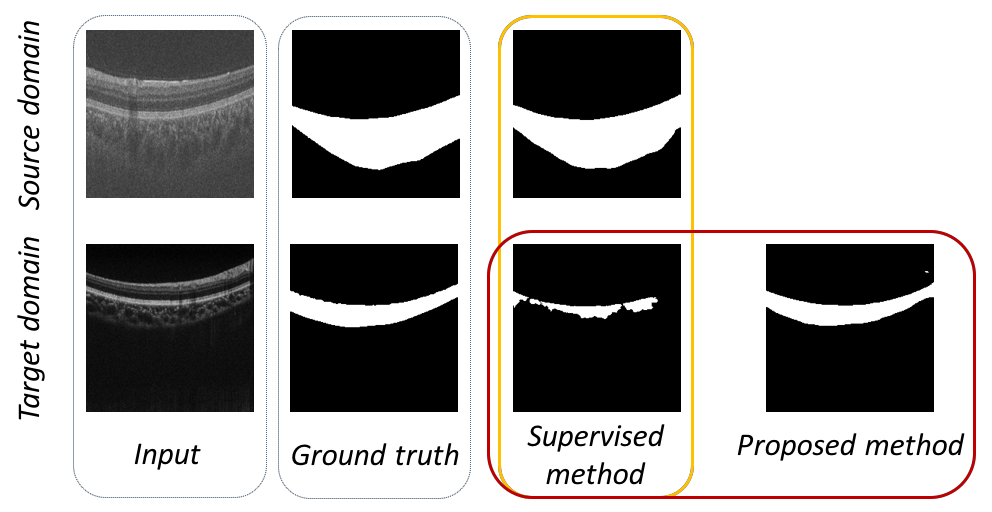}
	\caption{ The orange rounded rectangle shows that segmentation degrades due to domain discrepancy. The red rounded rectangle shows that domain adaptation method can obtain a better segmentation in target domain by minimizing domain discrepancy between the source and target domains.} 
	\label{fig:problem_show}
\end{figure}

\begin{figure*}[htb]
	\centering
	\includegraphics[width=6.8in]{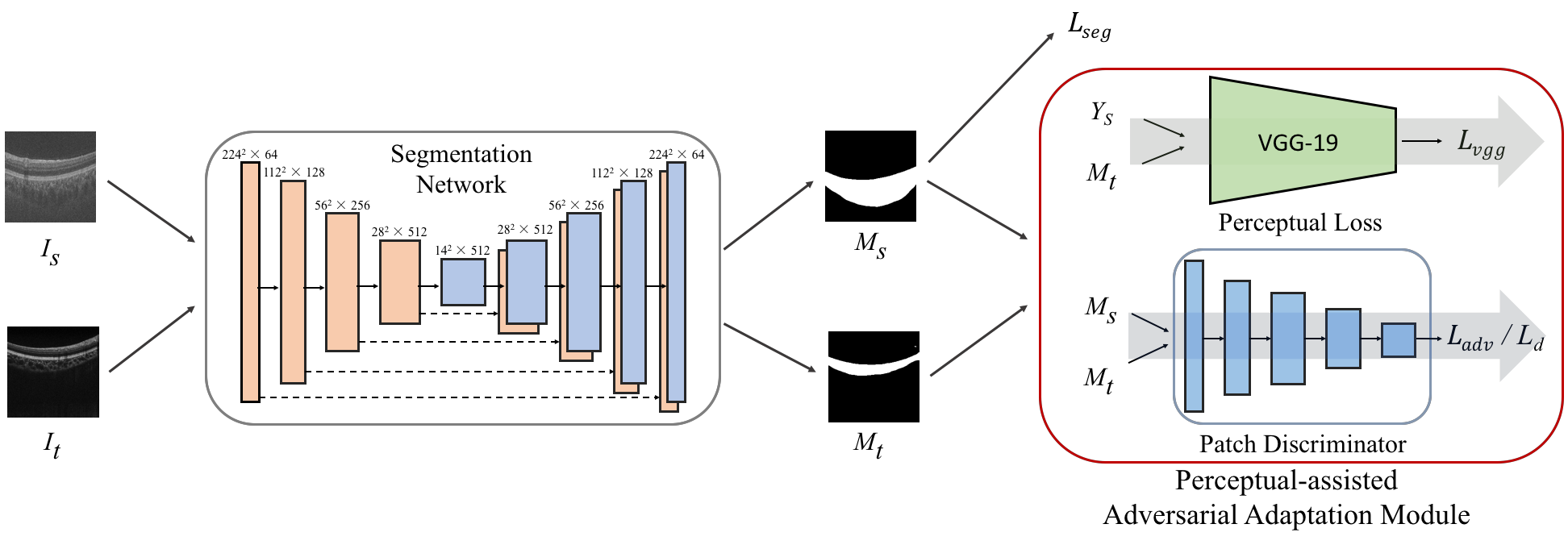}
	\caption{Overview of the Perceptual-assisted Adversarial Adaptation method. Firstly, source images ($I_{s}$) and target images ($I_{t}$) are fed into identical segmentation network based on U-Net~\cite{Olaf2015UNet} and obtain the predictions ($M_{s}$ and $M_{t}$) of source and target images, respectively. Then, the prediction of the source image ($M_{s}$) is utilized to compute segmentation loss. Finally, the predictions of the source image and the target image are fed into the perceptual-assisted adversarial adaptation (PAAA) module. In PAAA module, the prediction of the target image ($M_{t})$ and the label of the source image ($Y_{s}$) are fed into perceptual loss module while the predictions of the source image and the target image ($M_{s}$ and $M_{t}$) are fed into patch discriminator module. The whole network was trained in an end-to-end manner.} 
	\label{fig:architecture}
\end{figure*}

In recent years, deep learning has shown its advantage for medical image analysis~\cite{LITJENS201760}, especially the convolutional neural network (CNN) which is trained in an end to end manner has been widely used for medical image segmentation. Massod~\etal~\cite{Masood2019Automatic} and Caneiro~\etal~\cite{Caneiro2019Automatic} have shown the advantage of deep learning methods compared to non-deep learning methods in choroid segmentation task. In ~\cite{Masood2019Automatic}, the authors proposed a two-stage segmentation network with a combination of CNNs and morphological operations.  In ~\cite{Caneiro2019Automatic}, patch-based supervised machine learning methods were used to segment retinal and choroid boundary. The supervised methods based on deep learning have brought significant improvement for choroid segmentation. Although the supervised segmentation methods (\eg , FCN, U-Net) have a outstanding performance for choroid segmentation, there are two challenges for supervised methods. 1) supervised deep learning fails to obtain satisfactory segmentation results on new datasets due to the domain discrepancy between different datasets. 2) pixel-level annotations are time-consuming and labor-intensive in existing supervised methods for choroid segmentation. The orange rounded rectangle in Fig. ~\ref{fig:problem_show} show that segmentation degrades when the supervised models are adapted from the source domain to the target domain.

To overcome the shortcomings of supervised deep learning, an unsupervised domain adaptation method aimed at minimizing the domain shift (domain discrepancy) has been proposed over the past few years~\cite{wang2018deep}. Tsai~\etal~\cite{Tsai2018Learning} considered semantic segmentation as structured outputs and conduct adversarial learning in output space for semantic segmentation. Wang~\etal~\cite{Wang2019Patch} introduced an unsupervised domain adaptation method using a patch discriminator for joint optic disc and cup segmentation. Although the unsupervised domain adaptation methods mentioned above can obtain better performance when facing the domain discrepancy, adversarial adaptation can only narrow the prediction distribution between the source domain and the target domain. It is difficult for adversarial learning to capture details. In this paper, we present a perceptual-assisted adversarial adaptation for choroid segmentation. Our PAAA framework incorporates adversarial adaptation to address the domain discrepancy by encouraging the structure information of the target domain close to that of the source domain. Furthermore, we utilize the perceptual loss module to match shape information of the source and target domains which can result in a fine boundary prediction.

\textbf{Contributions:} 1) We propose a perceptual-assisted adversarial adaptation method to minimize the performance degradation caused by domain discrepancy. 2) To the best of our knowledge, our work is the first of its kind to leverage the combined advantage of adversarial adaptation and perceptual loss for choroid segmentation in OCT in an unsupervised manner. 3) We apply our method to segment choroid area in OCT data. Results show that our proposed method outperforms the state-of-the-art methods.  

\section{METHOLODY}
\label{sec:method}

\subsection{Problem Formulation}

As shown in Fig.~\ref{fig:architecture}, our proposed method includes two parts: the segmentation network $\mathbf{G}_{seg}$ and the perceptual-assisted adversarial adaptation part which combines patch discriminator $\mathbf{D}_{adv}$ module with perceptual loss module $\mathbf{P}$ to implement domain adaptation in output space. Formally, the source domain and target domain are denoted as $S$ and $T$, respectively. The input images from source domain are denoted as $\mathcal{I}_s$ $\in \mathbf{R}^{C \times W \times H}$ while the ones from target domain are denoted as $\mathcal{I}_t$ $\in \mathbf{R}^{C \times W \times H}$. $M_{s}$ and $M_{t}$ represent the predictions of the source image and the target image while $Y_{s}$ denotes the label of source image.

\subsection{Learning at Source Domain}

We first adopt the cross-entropy loss as the segmentation loss for images from the source domain to train the segmentation network:

\begin{equation}
\mathcal{L}_{seg}(\mathcal{I}_s) = - \sum_{h, w}\sum_{c \in C}{Y_s^{(c,h,w)}log(\mathcal{M}_s^{(c,h,w)})}
\end{equation}

where $Y_s$ is the ground truth annotaions for source images and $\mathcal{M}_s = \mathbf{G}_{seg}(\mathcal{I}_s)$  is the output of segmentation network. $\mathbf{G}_{seg}$ represents the segmentation network.

\subsection{Adversarial Adaptation for Target Domain}
The segmentation loss $\mathcal{L}_{seg}$ is only applied on the source domain. To optimized the network for images $\mathcal{I}_t$ in target domain, we forward them into $\mathbf{G}_{seg}$ to obtain the segmentation prediction $\mathcal{M}_t = \mathbf{G}_{seg}(\mathcal{I}_t)$. Then we first leverage the adversarial training to minimize the discrepancy between the prediction of the target domain and the one of the source domain in output space. Adversarial training is achieved by utilizing  a PatchGAN loss \cite{pix2pix2017}. Two predictions $\mathcal{M}_s$ and $\mathcal{M}_t$ generated from the network are fed into the discriminator $\mathbf{D}_{adv}$. To make the distribution of $\mathcal{M}_t$ closer to $\mathcal{M}_s$, we use an adversarial loss $\mathcal{L}_{adv}$ as :

\begin{equation}
\mathcal{L}_{adv}(\mathcal{I}_t) = -\sum_{h, w}log(\mathbf{D}_{adv}(\mathcal{M}_t)^{(c,h,w)})
\end{equation}

This loss is designed to train the segmentation network and fool the discriminator by maximizing the probability of the target prediction being considered as the source prediction. Through adversarial optimization, the output space of the target domain learns to mimic the distribution of the source output space.

\subsection{Perceptual-assisted Adaptation for Target Domain}

%In the initial training stage, the output of images in source domain are not convergent. So it is hard to training discriminator $\mathbf{D}_{adv}$. To address this issue,  we propose to transfer the mask style of target prediction $\mathcal{M}_t$ to source ground truth $\mathcal{Y}_s$ by utilizing the pre-trained VGG-19 network. 

Besides, a perceptual loss~\cite{Gatys_2016_CVPR} $\mathcal{L}_{per}$ is employed for matching their shape information (the curvatures of Bruch's membrane and choroid-sclera interface) which can result in a fine boundary prediction. As the name suggests, $\mathcal{L}_{per}$ penalizes results that are not perceptually similar to labels by defining a distance measure between feature maps of target prediction mask and source label after passing a pretrained VGG network. The constriction matches the shape information between the source domain and target domain which drives the prediction of the choroid area close to the true physiological structure and makes a better boundary prediction. Perceptual loss is defined as 

\begin{equation}
\mathcal{L}_{per}(\mathcal{I}_t) = \sum_i \dfrac{1}{N_i} || \phi_i(\mathcal{M}_t) - \phi_i(\mathcal{Y}_s)||_1
\end{equation}

where $\phi_i$ is the feature map of the i'-th layer of the pre-trained network. For this work, $\phi_i$ denotes the feature maps from layers \textit{relu1\_1, relu2\_1, relu3\_1, relu4\_1} and \textit{relu5\_1} of the VGG-19 \cite{Simonyan2015Very} network pretrained on the ImageNet \cite{Deng2009Imagenet} dataset.

The overall training objective for segmentation network is:

\begin{equation}
\mathcal{L} = \lambda_{seg} \mathcal{L}_{seg}(\mathcal{I}_s) + \lambda_{adv} \mathcal{L}_{adv}(\mathcal{I}_t) + \lambda_{per} \mathcal{L}_{per}(\mathcal{I}_t)
\end{equation}

By incorporating adversarial adaptation with perceptual loss based on VGG network, the proposed method can minimize the discrepancy between the source domain and the target domain in output space.
\section{Experiments}
\label{sec:exp}

\begin{figure}[htb]

\begin{minipage}[b]{0.14\linewidth}
  \centering
  \centerline{\includegraphics[width=1.4cm]{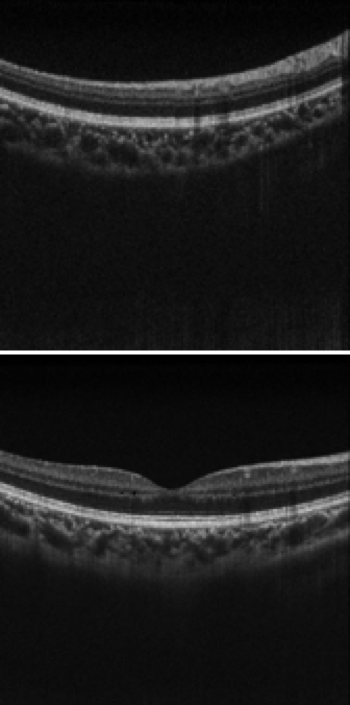}}
%  \vspace{2.0cm}
  \centerline{a}\medskip
\end{minipage}
\hfill
\begin{minipage}[b]{0.14\linewidth}
  \centering
  \centerline{\includegraphics[width=1.4cm]{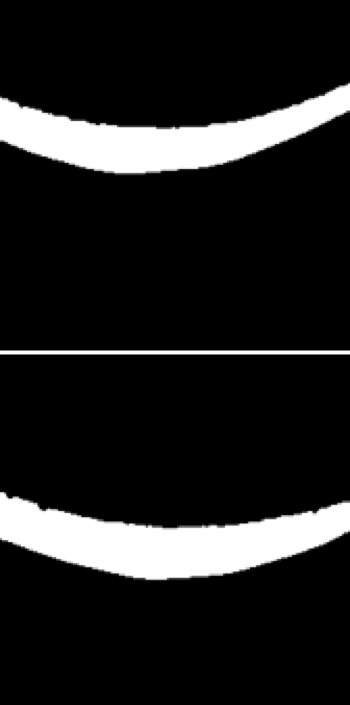}}
%  \vspace{1.5cm}
  \centerline{b}\medskip
\end{minipage}
\hfill
\begin{minipage}[b]{0.14\linewidth}
  \centering
  \centerline{\includegraphics[width=1.4cm]{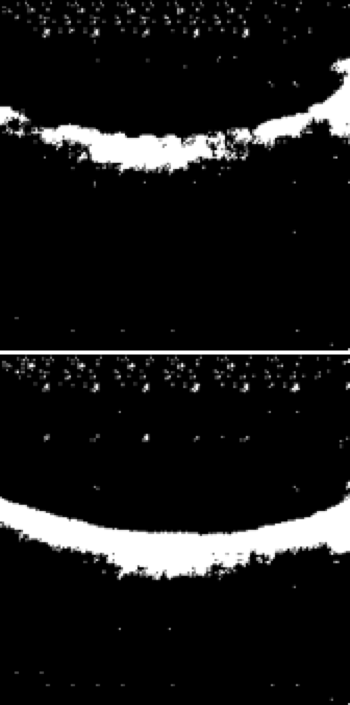}}
%  \vspace{1.5cm}
  \centerline{c}\medskip
\end{minipage}
\hfill
\begin{minipage}[b]{0.14\linewidth}
  \centering
  \centerline{\includegraphics[width=1.4cm]{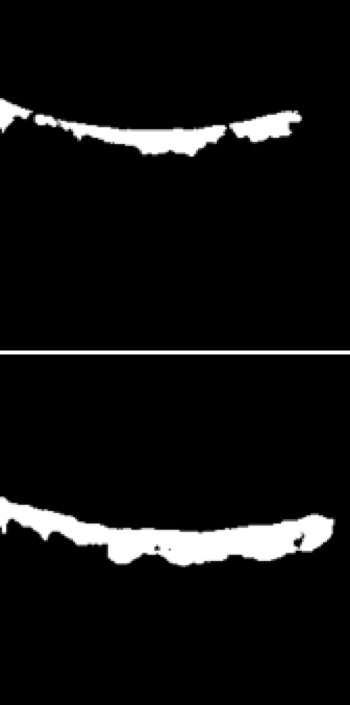}}
%  \vspace{1.5cm}
  \centerline{d}\medskip
\end{minipage}
\hfill
\begin{minipage}[b]{0.14\linewidth}
  \centering
  \centerline{\includegraphics[width=1.4cm]{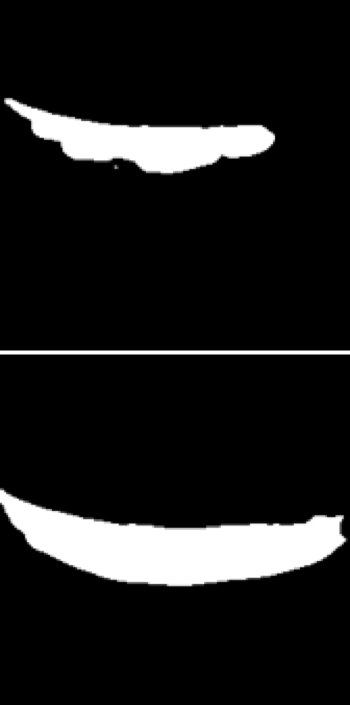}}
%  \vspace{1.5cm}
  \centerline{e}\medskip
\end{minipage}
\hfill
\begin{minipage}[b]{0.14\linewidth}
  \centering
  \centerline{\includegraphics[width=1.4cm]{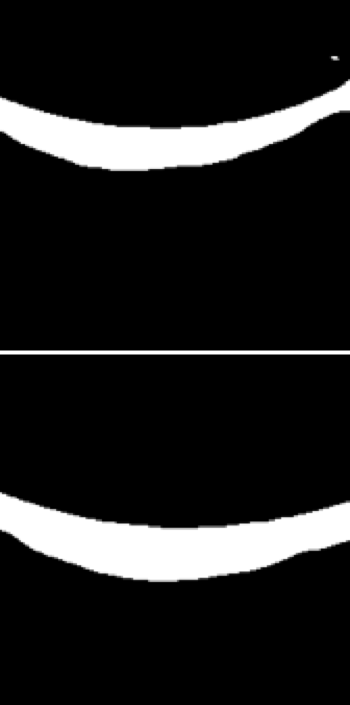}}
%  \vspace{1.5cm}
  \centerline{f}\medskip
\end{minipage}
\caption{Comparison of qualitative results when models are adapted from source domain image to target domain image. (a) Input. (b) Ground truth. (c) FCN. (d) U-Net. (e) AdaptSegNet. (f) Ours. }
\label{fig:results_show}
\end{figure}

\subsection{Datasets}

Two different datasets obtained from topcon OCT device and nidek OCT device are utilized to implement experiments in this paper. The two datasets are called TOPCON and NIDEK, respectively. Both of TOPCON and NIDEK datasets have pixel-level manual annotations. We set the train part of TOPCON dataset as the source domain and the train part of NIDEK dataset as the target domain. The train part of TOPCON dataset consists of 640 OCT images with 512 $\times$ 512 resolution and the train part of NIDEK dataset includes 120 OCT images with 512 $\times$ 1024 resolution. The test part of NIDEK dataset has 30 OCT images and are used for evaluation.

\subsection{Implementation details}
The architecture of the proposed method can be referred to Fig \ref{fig:architecture}. The proposed method leverage U-Net~\cite{Olaf2015UNet} as the backbone. The method was implemented in Python based on PyTorch. In the experiments, both of source image $\mathcal{I}_{s}$ and target image $\mathcal{I}_{t}$ were resized to 224 $\times$ 224 resolution. We trained the entire network end-to-end using Adam optimizer \cite{Diederik2015Adam}. The beta1 and beta2 of Adam optimizer was set to 0.9 and 0.99, respectively. The weight decay factor was 0.0001. The learning rate of the segmentation network was set to $10^{-3}$ and the one of the discriminator was set to 0.0001.  $\lambda_{seg}$, $\lambda_{adv}$, $\lambda_{per}$ in training objective were set to 100, 0.01, 0.06. Only the segmentation module was utilized in the evaluation phase.

\begin{table*}
\centering

\begin{threeparttable}

\begin{tabular}{ccccccc}
%\hline
%\multicolumn{7}{c}{TOPCON(source domain) --\textgreater NIDEK(target domain)}\\
\toprule
Methods & Source domain & Target domain & $AUSDE\tnote{1}\downarrow$ & $IOU\tnote{2}\uparrow$ & $GAP_{ausde}\downarrow$ & $GAP_{iou}\downarrow$\\
\hline
Oracle & NIDEK & NIDEK & 2.65 & 89.30 & - - - & - - -\\
\hline
FCN~\cite{Long2015FCN} & TOPCON & NIDEK & 53.08 & 42.97 & 50.43 & 46.33 \\
%\hline
U-Net~\cite{Olaf2015UNet} & TOPCON & NIDEK & 51.26 & 51.11 & 48.61 & 38.19 \\
%\hline
%CE-Net~\cite{Gu2019CENet} & TOPCON & NIDEK & 118.68 & 0.04 & 116.03 & 89.26 \\
%\hline
FCN + AA~\cite{Tsai2018Learning} & TOPCON & NIDEK & 24.50 & 51.19 & 16.98 & 28.57\\
\hline
U-Net + AA & TOPCON & NIDEK & 19.63 & 60.73 & 16.98 & 28.57\\
%\hline
%AdaptSegNet(U-Net) + FM & TOPCON & NIDEK & 4.36 & 83.44 & 1.71 & 5.86\\
%\hline
\textbf{Proposed method} & TOPCON & NIDEK & \textbf{3.21} & \textbf{85.77} & \textbf{0.56} & \textbf{3.53}\\
\bottomrule
\end{tabular}

\begin{tablenotes}
\footnotesize
\item[1] AUSDE (pixels)
\item[2] IOU ($\%$)
\end{tablenotes}

\end{threeparttable}

\caption{Results of adapting from the source domain to the target domain. We compare the results of our proposed method with other methods. }
\label{tab:results_compare}
\end{table*}

\subsection{Metrics}
\label{subsec:metrics}
In order to properly evaluate the performance of our proposed method, the following metrics will be utilized in experiments.

\textbf{Intersection over uniou (IOU):} Intersection over uniou is  a common evaluation metric to evaluate the segmentation quality. The IOU is defined as:  $ IOU = A_{o}$ / $A_{u} $, where $A_{o}$ and $A_{u}$ represent area of overlap and area of union between the prediction and the reference standard.
 
\textbf{Average unsigned surface detection error (AUSDE) \cite{Xiang2018Automatic}:} Average unsigned surface detection error was computed for each lower boundary of choroid by measuring absolute Euclidean distance in the z-axis between the results of the algorithm and the reference standard. AUSDE is efficient metric to evaluate performance of choroid segmentation between different algorithms.

Moreover,  $GAP$ ($GAP_{ausde}$ and $GAP_{iou}$) is another criterion to evaluate the adaptation performance of models. $GAP_{\mathcal{X}}$ is the gap of $\mathcal{X}$ between the model and the fully-supervised model, where $\mathcal{X}$ denotes ausde or iou.

%Another factor to evaluate the generalization ability of different methods is to measure the gap between the results of adapting from source domain to target domain and results of training with target domain at a supervised manner.

\subsection{Results and Discusions}
We evaluate the performance of our proposed method using the metrics mentioned in \ref{subsec:metrics} and compare our methods with other state-of-the-art methods when models are adapted from the source domain to the target domain.

%\textbf{Quantitative evaluation: }In the experiments, we evaluate our proposed method with TOPCON dataset and NIDEK dataset. we set TOPCON dataset as source domain and NIDEK dataset as target domain. In table~\ref{tab:results_compare}, the Oracle refer to the best result among those state of the art models training on target domain at a supervised manner and was set as the upper bound of segmentation performance of domain adaptation methods. We trained our proposed method using training set in source domain and training set (without label) in target domain. Other methods were trained with source domain. All the results in table~\ref{tab:results_compare} were tested with target domain. The table shows that the segmentation of other methods degrades due to domain discrepancy while our proposed method have a strong generalization for different domains. The results of our proposed method is close to the supervised one trained with target domain which requires time-consuming and labor-intensive pixel-level annotations.

\textbf{Quantitative evaluation: }In Table.~\ref{tab:results_compare}, we compare the proposed method with oracle (upper bound), supervised methods and unsupervised domain adaptation methods.  Except for oracle, all experiments take TOPCON as source domain and NIDEK as target domain. The oracle represents the results of U-Net trained and evaluated on the same target domain. The comparison results between the supervised method (U-Net) and oracle quantitatively show the performance degradation due to the domain discrepancy. Our method achieves over 48 pixels decreasing for AUSDE and over 34$\%$ IOU improvement compared with the supervised methods (FCN, U-Net), demonstrating the effectiveness of perceptual-assisted adversarial adaptation method used to narrow domain discrepancy. We also compare our method with the state-of-the-art unsupervised domain adaptation method AdaptSegNet (FCN + AA) and our method outperforms the AdaptSegNet by more than 21 pixels decreasing for AUSDE and more than 34$\%$ IOU improvement, which show that the proposed PAAA method can obtain a better segmentation performance than AdaptSegNet.  The ablation experiments were implemented, the proposed method can decrease the AUSDE by more than 16 pixels and improve the IOU by more than 25$\%$ compared with the baseline method (U-Net + AA) without perceptual loss module. The ablation experiments demonstrate that the perceptual loss help to result in a fine boundary prediction and obtain a better segmentation result. The $GAP_{ausde}$ and $GAP_{iou}$ indicate that the segmentation results of our proposed method are closer to that of oracle than other methods. what's more, the proposed method don not require pixel-level annotations in the target domain while the oracle needs the labor-intensive and time-consuming annotations to be trained. 

\textbf{Qualitative evaluation: }Fig.~\ref{fig:results_show} shows the visualization of choroid segmentation results when models are adapted from source domain image to target domain image. The predictions of our methods are visually better than that of other methods and more closer to the label (ground truth). The comparison shows that the proposed method has a better segmentation performance than other methods.

\section{Conclusion}
\label{sec:conclu}
In this paper, we have proposed a novel framework Perceptual-assisted Adversarial Adaptation to narrow the discrepancy between the source domain and the target domain and to obtain a better segmentation performance. The experiment results have demonstrated that our proposed method has an outstanding performance for choroid segmentation compared with other methods when models are adapted from the source domain to the target domain.

% Below is an example of how to insert images. Delete the ``\vspace'' line,
% uncomment the preceding line ``\centerline...'' and replace ``imageX.ps''
% with a suitable PostScript file name.
% -------------------------------------------------------------------------

\vfill
\pagebreak

\bibliographystyle{IEEEbib}
\bibliography{refs}

\end{document}